# The ESO Distant Cluster Sample: galaxy evolution and environment out to z=1

Bianca M. Poggianti[1], Alfonso Aragón-Salamanca[2], Steven Bamford[2], Fabio Barazza[3], Philip Best[4], Douglas Clowe[5], Julianne Dalcanton[6], Gabriella De Lucia[7], Vandana Desai[8], Rose Finn[9], Claire Halliday[1], Pascale Jablonka[3,10,11], Olivia Johnson[2], Bo Milvang-Jensen[12], John Moustakas[13], Stefan Noll[14], Nina Nowak[15], Roser Pelló[16], Sebastien Poirier[11], Gregory Rudnick[17], Roberto Saglia[15], Patricia Sánchez-Blázquez[18], Luc Simard[19], Jesus Varela[1], Anja von der Linden[20], Ian Whiley[2], Simon D.M. White[20], Dennis Zaritsky[21]

[1]INAF-Osservatorio Astronomico di Padova, IT
[2]University of Nottingham, UK
[3]Ecole Polytechnique Federale de Lausanne, CH
[4]Institute for Astronomy, Edimburgh, UK
[5]Ohio University, USA
[6]University of Washington, Seattle, USA
[7]INAF-Osservatorio Astronomico di Trieste, IT
[8]Spitzer Science Center, California Institute of Technology, USA
[9]Siena College, Loudonville, NY, USA
[10]Universite de Geneve, Observatoire, CH
[11]GEPI, Observatoire de Paris, Meudon, FR
[12]Dark Cosmology Centre, Niels Bohr Institute, University of Copenhagen, DK
[13]New York university, USA
[14]Observatoire Astronomique Marseille-Provence, FR
[15]Max-Planck-Institut fur extraterrestrische Physik, Garching, DE
[16]Universite de Toulouse, FR
[17]NOAO Tucson, USA
[18]University of Central Lancashire, UK
[19]Herzberg Institute of Astrophysics, Victoria, Canada
[20]Max-Planck-Institut fur Astrophysik, Garching, DE
[21]Steward Observatory, Tucson, USA

**The ESO Distant Cluster Survey (EDisCS, P.I. Simon D.M. White, LP 166.A-0162) is an ESO large programme aimed at studying clusters and cluster galaxies at z=0.4-1. How different is the evolution of the star formation activity in clusters, in groups and in the field? Does it depend on cluster mass and/or the local galaxy density? How relevant are starburst and post-starburst galaxies in the different environments? Is there an evolution in the galaxies' structures, and if so, is this related to the changes in their star formation activity? These are some of the main questions that have been investigated using the EDisCS dataset.**

There is no shortage of evidence that galaxy properties vary systematically with the environment in which they reside. The distributions of star formation histories, morphologies and masses, all strongly depend on galaxy location, at all redshifts probed so far. For large statistical studies, there are two main complementary ways to "measure" environment: the local galaxy number density (the number of galaxies per unit volume or projected area around the galaxy of interest), and the virial mass of the cluster or group where the galaxy resides, when this applies, usually derived from the velocity dispersion of the system.

Traditionally, studies of the dependence of galaxy evolution on environment were largely based on cluster studies, and their comparison with the "field". Recently, understanding the role of environment has become a major theme of all deep galaxy redshift surveys. Fully characterizing environment is a challenging task even for the largest field surveys, but great advances have been made in studying the local galaxy density as well as significant samples of groups at high-z. Even the widest area field surveys, however, include only a few distant massive systems. Only pointed surveys are able to probe galaxy clusters and massive groups, and to reach the high end of the local density distribution. Studies of galaxy clusters have traditionally opened the way to some of the major discoveries of galaxy evolution: to name a few, it is in clusters that astronomers first recognized the strong decline of the star formation activity in the last 7Gyr, the existence of a relation between local density and distribution of morphological types, the the fact that both star formation activity and mass assembly of massive ellipticals are completed at high-z, and the observed evolution of galaxy morphologies. Today, it has become possible to study galaxy evolution across a wide range of environments (clusters, groups, but also poor groups and the field), using optically selected cluster fields. Here we present one of such cluster surveys.

**The survey**

The ESO Distant Cluster Survey (hereafter, EDisCS) is a multiwavelength survey of galaxies in 20 fields containing galaxy clusters at z=0.4-1 based on an ESO Large Programme approved in P66.

Candidate clusters were chosen from among the brightest objects identified in the Las Campanas Distant Cluster Survey (Gonzales et al. 2001). They were confirmed by identifying red sequences in moderately-deep two-colour data from VLT/FORS2 (White et al. 2005). For all 20 fields, EDisCS assembled deep three-band optical photometry with FORS2/VLT (White et al. 2005, **Fig.1**), near-IR photometry in one or two bands with SOFI/NTT (Aragón-Salamanca et al. in prep.), deep multi-slit spectroscopy with FORS2/VLT (Halliday et al. 2004, Milvang-Jensen et al. 2008), and wide-field three-band imaging with MPG/ESO 2.2/WFI, as well as ACS/HST mosaic imaging of 10 of the highest redshift clusters (Desai et al. 2007). Other follow-up programmes have included XMM-Newton X-Ray observations (Johnson et al. 2006), Spitzer IRAC and MIPS imaging (Finn et al. in prep.), and Halpha narrow-band imaging (Finn et al. 2005). Photometric redshifts were derived by combining optical and near-IR imaging using two independent codes (Pello et al. 2009).

The EDisCS dataset has allowed the study of galaxies in a wide range of environments using homogeneous data. The EDisCS fields contain 16 clusters with velocity dispersion > 400 km/s, 10 groups with at least 8 spectroscopically-confirmed members for which a velocity dispersion between 150 and 400 km/s could be measured, as well as comparison samples of poor groups (galaxy associations of 3 to 6 galaxies) and of "field" galaxies not belonging to any cluster, group or poor group.
One of the most valuable characteristics of this dataset has proved to be the large range of velocity dispersions, and thus masses, of the cluster sample ( **Fig.2**). The quality of the cluster mass estimates based on VLT spectroscopy has been confirmed by weak lensing (Clowe et al. 2005) and X-ray (Johnson et al. 2006) estimates. EDisCS is thus the first distant cluster sample that can be used to assess how the evolution of galaxy properties depends on their host mass, providing a sample of high-z clusters that will evolve into a wide range of cluster masses today.

**Galaxy evolution: from star forming to passively evolving.....**

Early-type galaxies in clusters exhibit a well-defined relation between colour and luminosity, the so-called red sequence. Distinct red sequences have been observed for clusters up to z=1.5, indicating the existence of significant numbers of passively-evolving galaxies whose star formation activity terminated well before the observation epoch. However, *not all* today's red-sequence galaxies have been red and passive since high-z. There is solid evidence of a downsizing effect in the red-sequence build-up in clusters: on average, the most massive/luminous galaxies stopped forming stars, and therefore were already on the red sequence, at an earlier epoch than the less massive and less luminous ones. This has been demonstrated using EDisCS data in various independent ways. The evolution in the number ratio of luminous-to-faint red galaxies clearly shows a deficit of faint, red galaxies in distant clusters compared to local clusters (De Lucia et al. 2004, 2007, **Fig.3).** In agreement with this, the luminosity function of red galaxies is consistent with passive evolution at the bright end, but shows a significant build-up at the faint end towards lower redshifts (Rudnick et al. 2009, **Fig.3).** Relative to the number of bright red galaxies, the field has more faint red galaxies than clusters at z=0.6-0.8, but fewer at z=0.4-0.6. This variation in evolutionary rate with environment implies that cluster environments are more efficient than the field at adding galaxies to the red sequence at z<1. Finally, an analysis of spectral line indices yielding stellar population ages, metallicities and alpha-element enhancements confirms that massive red galaxies are well described by passive evolution and high formation redshifts, while less massive galaxies require a more extended star-formation history (Sánchez-Blázquez et al. 2009).

The most massive of all galaxies, the Brightest Cluster Galaxies (BCGs), are no exception to the downsizing effect: the evolution of their colour and rest-frame K-band luminosity is consistent with stellar populations formed at z>2 and subsequent passive evolution. In contrast with previous findings, BCGs in EDisCS clusters do not show a significant change in stellar mass since z=1 (Whiley et al. 2008). The clusters with large velocity dispersions, and therefore masses, tend to have brighter and more massive BCGs, but this dependency is weak: the stellar mass of the BCG changes only by 70% over a two-order-of-magnitude range in cluster mass.

**.....in different environments....**

By comparing the fraction of emission-line galaxies in EDisCS clusters with that in low-z clusters Poggianti et al. (2006) have studied in detail the evolution of the star-forming galaxy fraction as a function of cluster mass. At high-z, there is a broad anticorrelation between the star-forming fraction and the system velocity dispersion (**Fig.4**). Low-z clusters have much lower star-forming fractions than clusters at z=0.4-0.8 and, in contrast with the distant clusters, show a plateau for velocity dispersions above ~500 km/s, where the star-forming fraction does not vary systematically with velocity dispersion (**Fig.4)**. Thus, the evolution in the average proportion of star-forming galaxies is strongest in intermediate-mass systems, those with sigma ~ 500-600 km/s at z=0.

This evolution in the fraction of star-forming galaxies parallels the strong evolution observed in the galaxy morphological fractions. Visually classifying galaxies into Hubble types on the basis of our ACS/HST images, we have confirmed previous results—from the MORPHS team and others—finding lower S0 and higher spiral fractions in distant clusters compared to nearby ones, suggesting that a large number of distant spirals have turned into (some of the) present day S0s (Desai et al. 2007, **Fig.5**). In contrast, no evidence for evolution in the elliptical fraction is observed out to z=1. The comparison of the stellar population analysis and the morphologies also confirms previous evidence that the timescale for morphological evolution is longer than the timescale over which the star formation activity ceases (Sánchez-Blázquez et al. 2009). EDisCS results demonstrate that cluster morphological fractions plateau beyond z~0.4, implying that most of the morphological evolution in luminous galaxies occurs in the last 5 Gyr (Desai et al. 2007).

Combined with other distant cluster samples, the EDisCS dataset shows a correlation between morphological content and cluster velocity dispersion similar to the relation between star forming fraction and velocity dispersion described above (Desai et al. 2007). Altogether, these results demonstrate that morphological evolution at z<1 involves a significant fraction of cluster disk galaxies, and it is stronger in lower mass clusters than in the most massive ones, as it is for the changes in star-forming fractions (Poggianti et al. 2009b). Thus, the strongest evolution between z~1 and today appears to affect galaxies in intermediate mass environments, those commonly named low-mass clusters or massive groups.

Disk-dominated galaxies in the EDisCS dataset are also found to have a higher fraction of large-scale bars than bulge-dominated galaxies, in agreement with local studies (Barazza et al. 2009). Samples of disk galaxies selected in clusters and in the field both show similar bar fractions and bar properties, indicating that internal processes are crucial for bar formation. On the other hand, cluster centres are found to be favourable locations for bars, which may suggest that the internal processes responsible for the bar growth are supported by the kind of interactions often taking place in these environments.

Analyzing the proportions of morphological types as a function of local galaxy density, EDisCS galaxies are found to follow a morphology-density relation similar to that observed in previous cluster studies, with ellipticals being more common in high density regions and spirals being more frequent in low density ones. The decline of the spiral fraction with density is entirely driven by galaxies of types Sc or later, while early spirals (Sa's and Sb's) are equally frequent at all densities, similarly to S0s (Poggianti et al. 2008). In addition to measuring the morphology-density relation, we have been able for the first time to quantify the relation between star-formation activity and local density in distant clusters. The star formation-density relation in EDisCS clusters qualitatively resembles that observed at low-z. In both nearby and distant clusters, higher-density regions contain proportionally fewer star-forming galaxies. This is observed using both the [OII] and the Hα lines (Poggianti et al. 2008, Finn et al. 2005). Moreover, the average [OII] equivalent width of star-forming galaxies is independent of local density both at high- and low-z. Although this seems to suggest that the star formation in star-forming galaxies is not affected by the local environment, that is not the case: the current average star formation rate in star-forming EDisCS galaxies *does depend* on the local density, and seems to peak at intermediate densities (15-40 galaxies/Mpc^2), declining at higher densities and also, possibly, at lower densities. So far, no one has studied how the star formation activity *in star-forming galaxies* varies with local density at low-z, thus our findings await a local comparison.

Finally, for galaxies of a given Hubble type, we see no evidence that star-formation properties depend on local environment. In contrast with recent findings at low-z, the star formation-density relation and the morphology-density relation in our distant clusters seem to be fully equivalent, suggesting that neither of the two relations is more fundamental than the other at these redshifts and/or in these environments (Poggianti et al. 2008).

### ...through a starburst and then a post-starburst phase?

A crucial question is, of course, on what timescale do galaxies turn from blue to red in different environments, as the timescale provides important clues on the physical processes involved.

The spectra of post-starburst galaxies (called "E+A" from their appearance of an elliptical galaxy spectrum with the strong Balmer lines of A type stars, or called "k+a" from their mixture of K stars and A stars) are characterized by their exceptionally strong Balmer lines in absorption and the lack of emission lines, belong to galaxies in which the star formation activity ended *abruptly* sometime during the past Gyr. In EDisCS we find that the incidence of k+a galaxies at these redshifts depends strongly on environment (Poggianti et al. 2009a). K+a's reside preferentially in clusters and, unexpectedly, in a subset of those groups that have a low fraction of [OII] emitters **(Fig.6)**. In these environments, 20-30% of the star-forming galaxies have had their star formation activity recently and suddenly truncated. In contrast, there are proportionally fewer k+a galaxies in the field, the poor groups, and groups with a high [OII] fraction.

The properties of k+a galaxies are consistent with previous suggestions that cluster k+a's are observed in a transition phase: at the moment they are rather massive S0 and Sa galaxies, evolving from star-forming, recently-infallen, later types to passively-evolving cluster early-type galaxies. The incidence of k+a galaxies correlates with the cluster velocity dispersion: more massive clusters have a higher proportion of k+a's. The correlation between k+a fraction and cluster velocity dispersion supports the hypothesis that k+a galaxies in clusters originate from processes related to the intracluster medium, while the origin of the high k+a frequency in low-[OII] groups is currently an important, but poorly understood, piece of the puzzle.

Spectra of dusty starburst candidates, with strong Balmer absorption and emission lines, present a very different environmental dependence from the post-starburst galaxies. They are numerous in all environments at z=0.4-0.8, but they are especially numerous in all types of groups, favoring the hypothesis of being triggered by a merger or tidal interaction (**Fig.6**). Hence, at least from the optical point of view, starbursts do not appear to be triggered by the cluster environment, while they probably do feed the cluster post-starburst population after they have fallen onto the clusters.

**Group bimodality?**

One of the unexpected and perhaps more striking results emerging from the EDisCS data is that two types of groups can be identified, with notably distinct galaxy properties but similar velocity dispersion estimates (100-400 km/s). There are groups with a star-forming fraction higher than 80% (high-[OII] groups), which consist mostly of morphologically late-type galaxies, with no post-starburst galaxies, thus quite similar to the "field" in all their galaxy properties. The other type of groups (low-[OII] groups) have low star-forming fractions (lower than 50%), consist mostly of galaxies with early-type morphologies, and have high proportions of post-starburst galaxies, i.e. their galaxy populations resemble those of the most massive clusters.

It can be argued that this dichotomy may arise simply because the low-[OII] groups are 'true" virialized groups, while the high-[OII] groups are just groups in formation or, more generally, unbound galaxy associations. However, this hypothesis is unlikely to explain the observed bimodality since a very large spread in star-forming fraction is also observed among X-ray selected groups, all of which have a hot intragroup medium within a potential well. The very different galaxy populations observed in the two types of groups could be the key to discriminate which physical processes establish the dependence of galaxy properties on environment. The answer will be found in future large (cluster or field) surveys with high-quality data.

Barazza, F., et al., 2009, A&A, 497, 713
Clowe, D., et al., 2006, A&A, 451, 395
De Lucia, G., et al. 2004, ApJL, 610, L77
De Lucia, G., et al. 2007, MNRAS, 374, 809
Desai, V., et al., 2007, ApJ, 660, 1151
Finn, R., et al., 2005, ApJ, 630, 206
Gonzalez, A.H., et al., 2001, ApJS, 137, 117
Halliday, C. et al., 2004, A&A, 427, 397
Johnson, O. et al., 2006, MNRAS, 371, 1777
Milvang-Jensen, B., et al., 2008, A&A, 482, 419
Pello, R., et al., 2009, A&A in press
Poggianti, B.M., et al., 2006, ApJ, 642, 188
Poggianti, B.M. et al., 2008, ApJ, 684, 888
Poggianti, B.M., et al., 2009a, ApJ, 693, 112
Poggianti, B.M., et al., 2009b, ApJL, in press (arXiv:0903.0504)
Rudnick, G., et al., 2009, A&A in press
Sánchez-Blázquez, P., et al., 2009, A&A in press (arXiv:0902.3392)
Whiley, I., et al., 2008, MNRAS, 387, 1253
White, S. et al., 2005, A&A, 444, 36

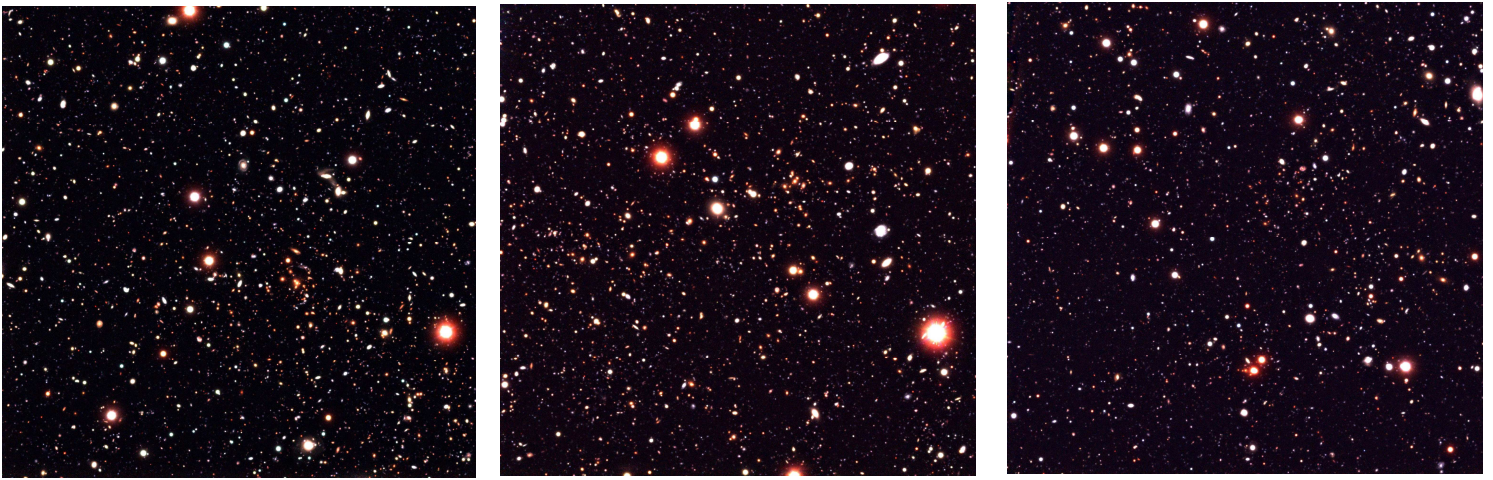

**Fig.1** Examples of EDisCS clusters: colour composite (VRI) FORS2 images of Cl1216 (z=0.79, velocity dispersion σ=1018 km/s, left), Cl1054-11 (z=0.70, σ=589 km/s, centre), Cl1103 (z=0.96, 0.70, 0.63, σ=534, 252, 336 km/s, right)

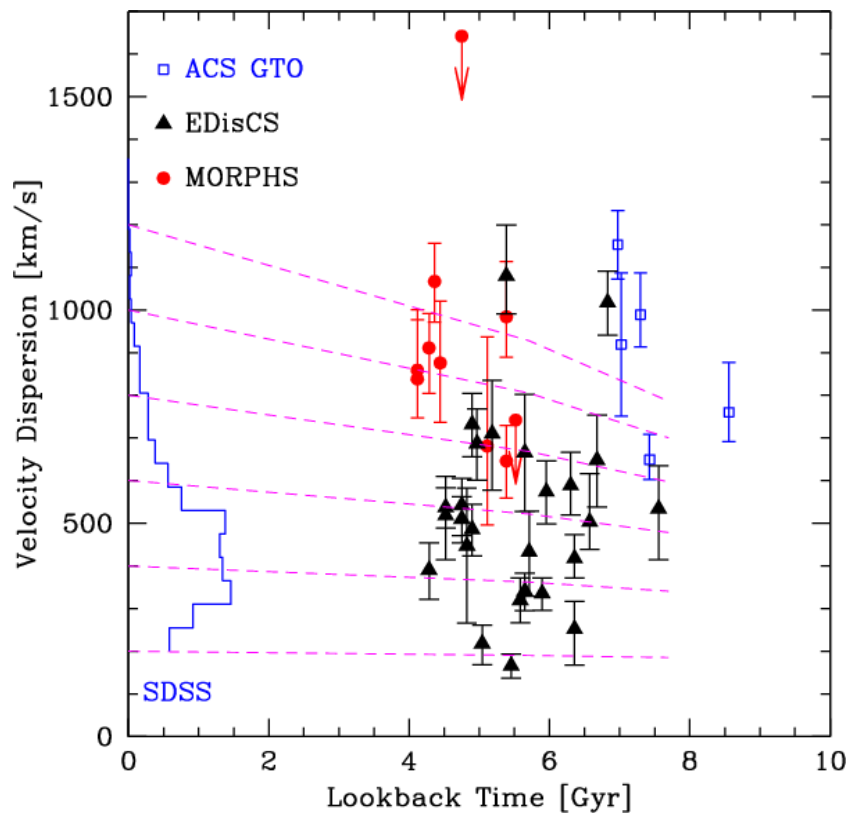

**Fig.2** The distribution of velocity dispersion vs. lookback time for EDisCS (black points) and for two other well-studied cluster samples at similar redshifts (red and blue points), as well as for a well-studied local sample (histogram). Dashed lines show how the velocity dispersion is expected to evolve with time. From this plot it is apparent that a) EDisCS clusters span a wide range of velocity dispersions/masses, and b) the majority of EDisCS clusters can be progenitors of "typical" low redshift clusters (Milvang-Jensen et al. 2008).

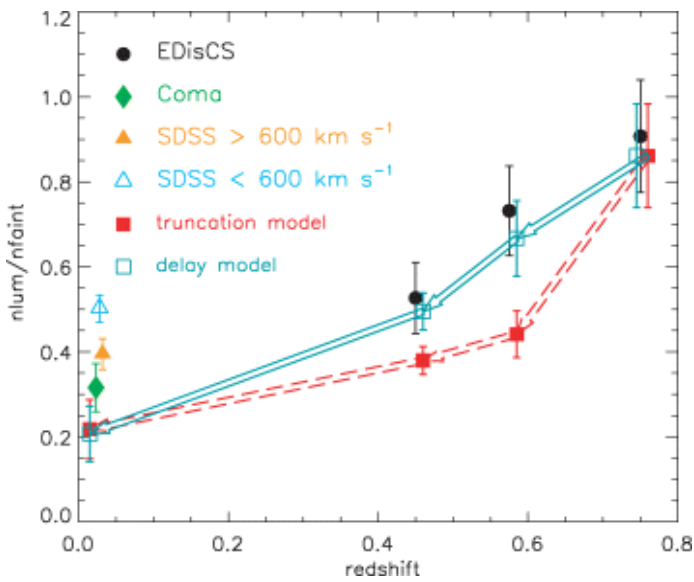
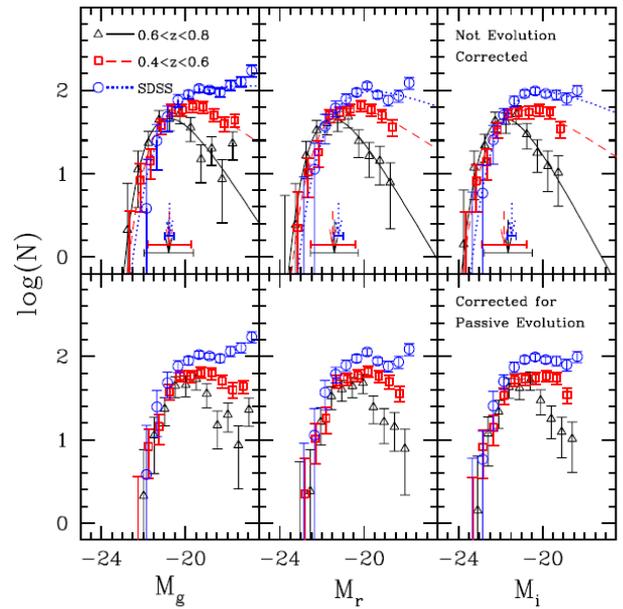

**Fig.3 Left:** Number ratio of luminous-to-faint red-sequence galaxies as a function of redshift. EDisCS results (black points) are compared with Coma and SDSS clusters at low-z. The arrows indicate the expected evolution if star formation is instantaneously or slowly truncated in star-forming galaxies infalling onto clusters (De Lucia et al. 2007) **Right:** Composite rest-frame $g$, $r$ and $i$ luminosity functions of red-sequence EDisCS and SDSS cluster galaxies. Luminosities are corrected (bottom panels) or uncorrected (top panels) for passive evolution (Rudnick et al. 2009).

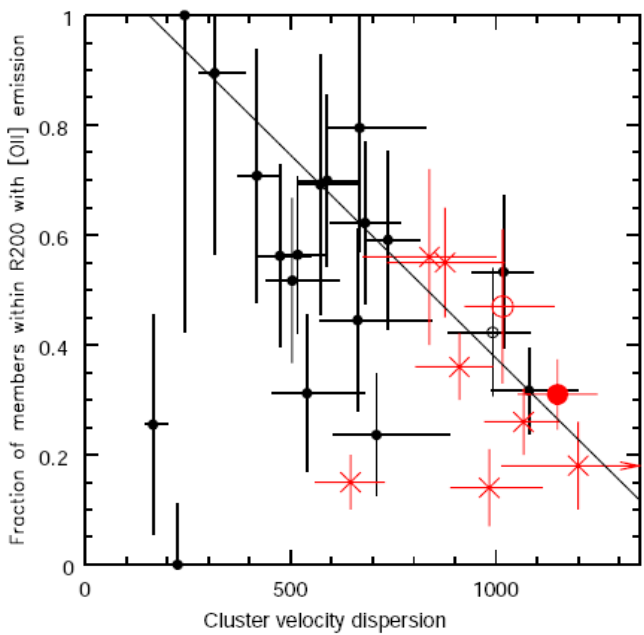
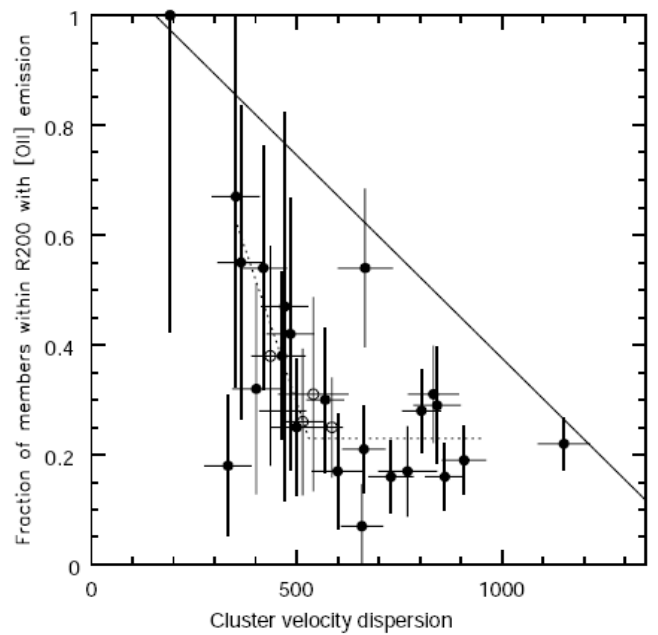

**Fig.4** Fraction of star forming galaxies vs cluster velocity dispersion in EDisCS clusters at z=0.4-0.8 (left) and in a cluster sample selected from Sloan at z=0.04-0.08 (right) (Poggianti et al. 2006). In the left panel, black points are EDisCS structures, red points are data from the literature. The solid line in both panels show the best fit to the high-z datapoints.

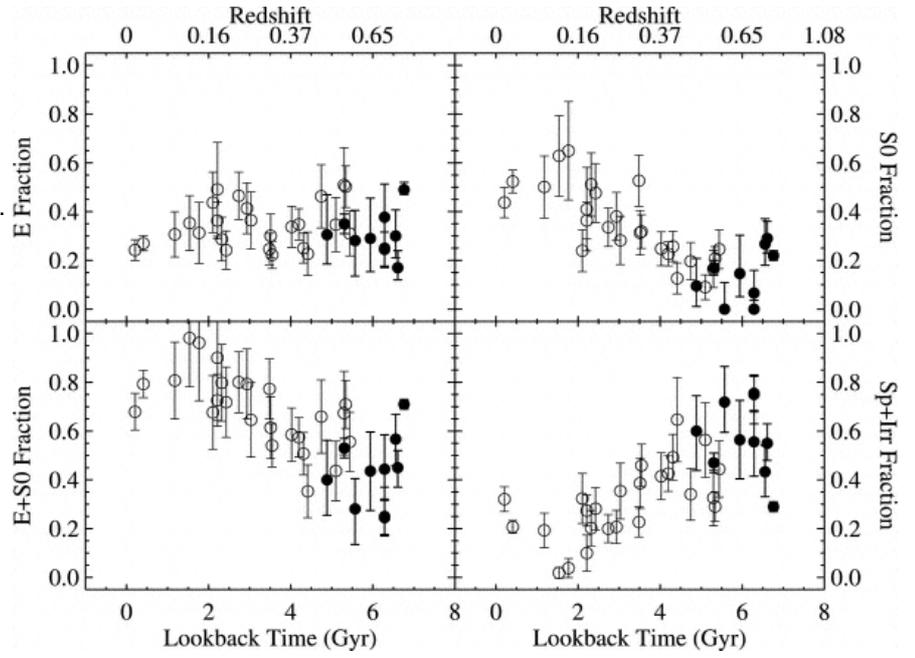

**Fig.5** Evolution of the elliptical, S0, early-type (E+S0) and spiral fractions (black points EDisCS, empty points from the literature) (Desai et al. 2007).

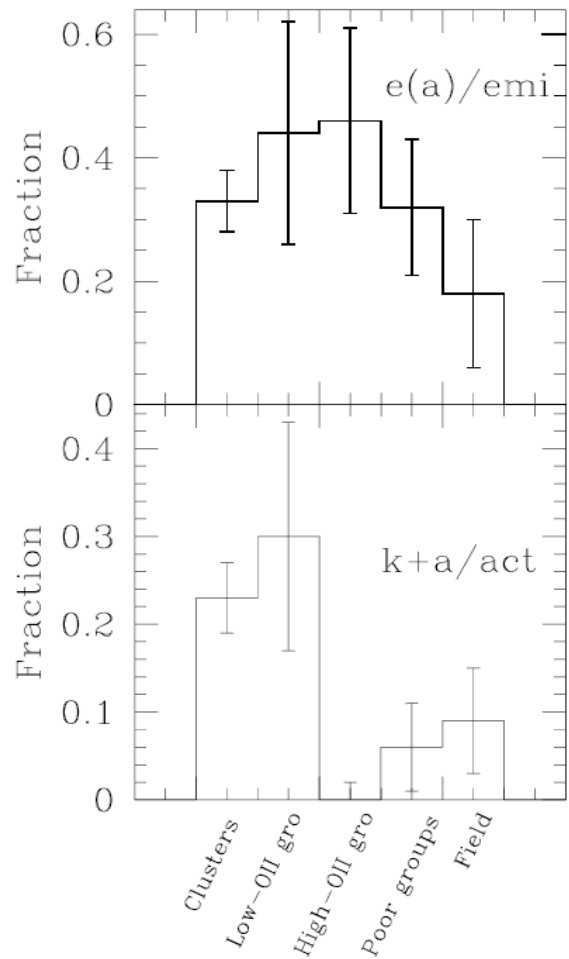

**Fig.6 Bottom.** Fraction of post-starburst galaxies among all galaxies with recent or ongoing star formation activity, in the different environments: clusters, groups with a low star-forming fraction (low-[OII] groups), groups with a high star-forming fraction (high-[OII] groups, poor groups and the field. **Top.** Fraction of candidate dusty starburst galaxies among all emission-line galaxies as a function of environment (Poggianti et al. 2009a).